\documentclass[onecolumn,prc,superscriptaddress,unsortedaddress,preprintnumbers,amsmath,amssymb]{revtex4}
\usepackage{graphicx}
\usepackage{dcolumn}
\usepackage{bm}
\usepackage{CJK}
\usepackage{CJKulem}
\usepackage{txfonts}
\usepackage[dvipdfmx,bookmarks=true,colorlinks,%
citecolor=blue,linkcolor=blue,anchorcolor=blue,filecolor=blue,urlcolor=blue,%
           ]{hyperref}

\usepackage{mathptmx, courier, pifont}
\usepackage[scaled=0.92]{helvet}
\usepackage[T1]{fontenc}
\usepackage{textcomp}
\usepackage{color}
\usepackage{colordvi}

\def\beq{\begin{equation}}
\def\eeq{\end{equation}}
\def\bea{\begin{eqnarray}}
\def\eea{\end{eqnarray}}

\def\fun#1#2{\lower3.6pt\vbox{\baselineskip0pt\lineskip.9pt
  \ialign{$\mathsurround=0pt#1\hfil##\hfil$\crcr#2\crcr\sim\crcr}}}

\begin{document}
\preprint{}
\title{Role of nucleon-nucleon correlation in transport coefficients and
gravitational-wave-driven $r$-mode instability of neutron stars}

\author{X. L. Shang}\affiliation{Institute of Modern Physics, Chinese
Academy of Sciences, Lanzhou 730000, China}\affiliation{School of
Physics, University of Chinese Academy of Sciences, Beijing 100049,
China}

\author{P. Wang}\affiliation{National Astronomical Observatories, Chinese Academy of
Sciences, Beijing 100012, China}

\author{W. Zuo}\affiliation{Institute of Modern Physics, Chinese
Academy of Sciences, Lanzhou 730000, China}\affiliation{School of
Physics, University of Chinese Academy of Sciences, Beijing 100049,
China}

\author{J. M. Dong}\email[ ]{dongjm07@impcas.ac.cn}\affiliation{Institute of Modern Physics, Chinese
Academy of Sciences, Lanzhou 730000, China} \affiliation{School of
Physics, University of Chinese Academy of Sciences, Beijing 100049,
China}

\date{\today}

\begin{abstract}
The thermal conductivity and shear viscosity of dense nuclear
matter, along with the corresponding shear viscosity timescale of
canonical neutron stars (NSs), are investigated, where the effect of
Fermi surface depletion (i.e., the $Z$-factor effect) induced by the
nucleon-nucleon correlation are taken into account. The factors which
are responsible for the transport coefficients, including the
equation of state for building the stellar structure, nucleon
effective masses, in-medium cross sections, and the $Z$-factor at
Fermi surfaces, are all calculated in the framework of the Brueckner
theory. The Fermi surface depletion is found to enhance the
transport coefficients by several times at high densities, which is more favorable to damping the
gravitational-wave-driven $r$-mode instability of NSs. Yet, the
onset of the $Z$-factor-quenched neutron triplet superfluidity
provides the opposite effects, which can be much more significant
than the above mentioned $Z$-factor effect itself. Therefore,
different from the previous understanding, the nucleon shear
viscosity is still smaller than the lepton one in the superfluid NS
matter at low temperatures. Accordingly, the shear viscosity cannot
stablize canonical NSs against $r$-mode oscillations even at quite
low core temperatures $10^6$ K.
\end{abstract}
\maketitle

As a class of compact objects, neutron stars (NSs) with
typical mass $M\sim 1.4 M_{\odot }$ and radii $R\sim 10$ km, contain
extreme neutron-rich matter at supranuclear density in their
interiors. Interestingly, they have many extreme features that
cannot be produced in terrestrial laboratories, such as extremely
strong magnetic field, superstrong gravitational field, extremely
high density, superfluid matter and superprecise spin
period~\cite{HPY}, suggesting their importance for fundamental
physics. These intriguing features have drawn great interest for
researchers of various branches of contemporary physics and
astronomy since the discovery of pulsars (rapidly rotating NSs) in
1967.

Due to the dense matter with large isospin asymmetry inside NSs, a
great deal of attention has been paid to the recent astronomical
observations that can be used to uncover the knowledge of the NS
interior. For instance, the observations of stellar cooling enables
one to constrain the equation of state (EOS) of dense matter,
superfluidity and transport properties, in combination with
indispensable theoretical
analysis~\cite{CAS1,Page1,CAS2,CAS3,CAS4,CAS5,CAS6}. Moreover, a
rapidly rotating NS is regarded as a gravitational wave source due
to $r$-mode instability. The $r$-mode is a non-radial oscillation
mode with Coriolis force as restoring force, which leads to the
gravitational wave radiation in rapidly rotating NSs due to the
Chandrasekhar-Friedmann-Schutz instability~\cite{CFS1,CFS2,CFS3} and
thus prevents the NSs from reaching their Kepler rotational
frequency~\cite{Kep1,Kep2}. The gravitational radiation is in
turn able to excite $r$ modes in NS core and hence enhances their
oscillation amplitudes, and it is particularly interesting from the
perspective of the gravitational wave observations with ground-based
facilities. The gravitational wave signal from the $r$-mode
oscillation, if detectable in the future, could help one to probe
the dense matter properties inside NSs.

The reliable knowledge about transport coefficients of dense matter
is crucial for understanding the stellar thermal evolution and
$r$-mode-instability induced gravitational radiation. The thermal
conductivity which measures the ability to conduct the heat, is an
important input for modeling NS cooling~\cite{Cool1,Cool2}. The
shear viscosity is the primary damping mechanism that hinders the
gravitational-wave-driven $r$-mode instability of rapidly rotating
NSs at low temperatures ($< 10^9$ K)~\cite{FI1979,CL1987,IV2012}.
These two transport coefficients have been calculated by several
authors based on the formulism derived by Abrikosov and Khalatnikov
(AK) from the Landau kinetic equations for a multicomponent
systems~\cite{AK}, where the required in-medium nucleon-nucleon
cross sections is obtained by employing the correlated basis
function method and the Brueckner-Hartree-Fock (BHF) approach with
realistic nucleon-nucleon
interactions~\cite{Benhar2007,Benhar2010,Zhang2012,Baldo2013}. In
the present work, within the AK framework, we calculate the
transport coefficients by adopting the Brueckner theory with the
inclusion of the effect of Fermi surface depletion. The bulk
viscosity is expected to become the dominant dissipation mechanism
for newborn NSs with rather high temperatures ($T>10^{10}$ K), and
we do not consider this situation here.

It is well-known that, the momentum distribution for a perfect Fermi
gas follows a right-angle distribution at zero-temperature, namely
the well-known Fermi-Dirac distribution. Yet, owing to the
short-range repulsive core and tensor interaction (collectively
referred to as short-range correlation in some references), the
system deviates from the typical profile of an ideal degenerate
Fermi gas featured by a high-momentum tail~\cite{SRC11,SRC12,SRC13,SRC14}, and as a result a Fermi
surface depletion may appear. The $Z$-factor
measures such a Fermi surface depletion. The correlation between nucleons
or its induced $Z$-factor has far-reaching impact on many issues
such as nuclear structure~\cite{Science2008,Science2014},
superfluidity of dense nuclear
matter~\cite{Dong-SRC1,Dong-SRC2,BAL}, NS cooling~\cite{Dong-SRC2}
and the European Muon Collaboration effect~\cite{Nature2018,EMC2},
highlighting its fundamental importance in nuclear physics and NS
physics. For instance, Dong et al. have shown that the superfluid
gap of $\beta$-stable neutron star matter is strongly quenched by
the $Z$ factor within the generalized BCS
theory~\cite{Dong-SRC1,Dong-SRC2}. The neutrino emissivity for NS
cooling due to direct Urca, modified Urca processes are also reduced
by the $Z$-factor, and therefore the cooling rates of young NSs are
considerably slowed~\cite{Dong-SRC2}.

\begin{figure}[htbp]
\begin{center}
\includegraphics[width=0.4\textwidth]{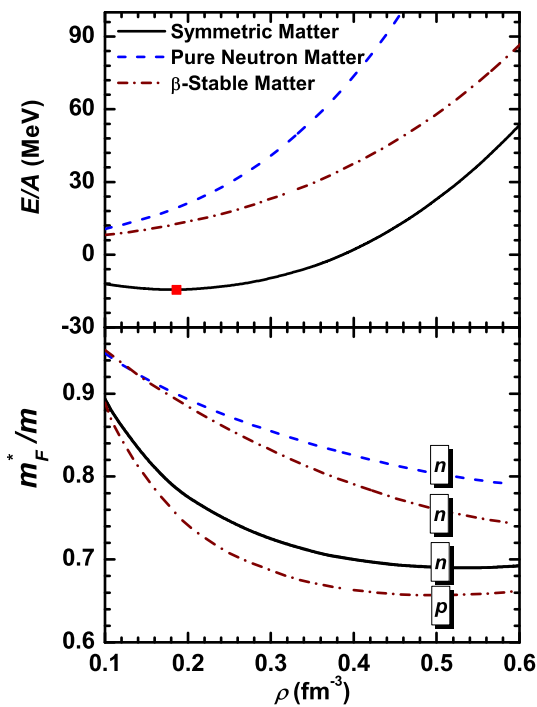}
\caption{(a) Energy per particle in symmetric matter, pure neutron
matter, and $\beta$-stable matter as a function of nucleonic density
from the BHF approach. The square shows the position of calculated
saturation point. (b) Density-dependent effective mass at Fermi
surfaces for three different nuclear matter
configurations.}\label{fig:EOS}
\end{center}
\end{figure}

In this work, the roles of the $Z$-factor in the thermal
conductivity and shear viscosity are clarified based on the AK
formulism. The neutron triplet superfluidity in NS core quenched by
the $Z$-factor effect is introduced to examine its effects on the
viscosity of $\beta$-stable NS matter. Then we calculate the shear
viscosity timescale and gravitation-wave-driven $r$-mode growth
timescale of canonical NSs to explore whether the shear viscosity is
sufficiently strong to damp the $r$-mode instability. The required
in-medium cross sections and nucleon effective masses to calculate
transport coefficients, and the $Z$-factor at the Fermi surface,
together with the EOS to establish the NS structure, are all
obtained in an unified framework, i.e., the Brueckner theory with
AV18 two-body interaction plus a microscopic three-body force
\cite{baldo,zuo}. We should stress here that in the calculation the
exact treatment of total momentum is adopted to obtain more reliable
results~\cite{shangbhf}.

The $Z$-factor that measures the effect of Fermi surface depletion
is given by
\begin{equation}
Z(k) = \left[
1-\frac{\partial\Sigma(k,\omega)}{\partial\omega}\right]_{\omega=\epsilon(k)}^{-1}
\end{equation}
with the single-particle energy $\epsilon(k)$.
Where $\Sigma(k,\omega)$ is the self-energy versus momentum $k$ and
energy $\omega$.  The $Z$ factor at the Fermi surface, labeled $Z_F$
($0<Z_F<1$), is equal to the discontinuity of the occupation number
at the Fermi surface, according to the Migdal-Luttinger
theorem~\cite{Migdal1960}. Once the nucleon-nucleon
correlation is included, the nucleon momentum distribution is given
as
\begin{eqnarray}
n(k)=\int\frac{d\omega}{2\pi}S(k,\omega)n^{0}(\omega)
\end{eqnarray}
at finite temperature $T$~\cite{KG1962}, where $\omega$ is the
energy.
$n^{0}(\omega)=1/[1+\exp(\frac{\omega-\mu}{k_{B}T})]$ is
the well-known Fermi-Dirac distribution function under temperature $T$ and chemical potential $\mu$.
The spectral function
$S(k,\omega)$ can be expressed as~\cite{baldo}
\begin{eqnarray}
S(k,\omega)\approx Z_{F}\delta(\omega-\epsilon(k_{F})),k\approx k_{F},
\end{eqnarray}
when momentum $k$ is extremely close to the Fermi momentum  $k_F$.
Consequently, the momentum distribution near the Fermi surface is
approximated by \cite{Dong-SRC2}
\begin{eqnarray}
n(x)\approx Z_{F}n^{0}(x),k\approx k_{F},
\end{eqnarray}
with $x=(\epsilon(k)-\mu)/(k_{B}T)$. Hereafter we take $x$ as
variable in the Fermi-Dirac distribution for convenience. We stress
that this approximation is only valid when $k$ is extremely close to
the Fermi surface. The nucleon-nucleon correlation quenches the
occupation probability by a factor $Z_F$ at Fermi surface $k_F$, and
thus it hinders particle transitions around the Fermi surface.

To embody the effects of nucleonic Fermi surface depletion in the
calculation of the kinetic coefficients, we extend the Landau
kinetic equation by including the $Z$-factor in the collision
integral. In the AK framework, at temperature $T$, the collision
integral without the $Z$-factor effect takes the form of~\cite{PRB35}
\begin{eqnarray}
I_{1i}^{0}&=&-\frac{m_{i}^{*}k_{B}^{2}T^{2}}{8\pi^{4}\hbar^{6}}\int\int
dx_{2}dx_{3}n^{0}(x_{1})n^{0}(x_{2})[1-n^{0}(x_{3})]\nonumber\\
&\times&[1-n^{0}(x_{1}+x_{2}-x_{3})]\sum_{j}m_{j}^{*2}\int\int
\frac{d\Omega}{4\pi}\frac{d\phi_{2}}{2\pi}\nonumber\\
&\times&\frac{W_{ij}(\theta,\phi)\beta_{ij}}{1+\delta_{ij}}[\psi(\bm{p}_{1})+\psi(\bm{p}_{2})-\psi(\bm{p}_{3})-\psi(\bm{p}_{4})],
\end{eqnarray}
where $m^{*}$ is the effective mass of nucleon $i$ or $j$. And the
small quantities $\psi(\bm{p})$ measures the departure from
equilibrium state. Here the nucleon-nucleon scattering is limited to
the Fermi surface. For convenience, one can assume $1$ and $3$ ($2$
and $4$) are the same component, i.e.,
$|\bm{p}_{1}|=|\bm{p}_{3}|=p_{i}$
($|\bm{p}_{2}|=|\bm{p}_{4}|=p_{j}$). And the transition probability
$W_{ij}$ from two quasiparticle state
$|\bm{p}_{1},\bm{p}_{2}\rangle$ to state
$|\bm{p}_{3},\bm{p}_{4}\rangle$, depends only on $\theta$ and $\phi$
($d\Omega=\sin\theta d\theta d\phi$), where $\theta$ is the angle
between $\bm{p}_{1}$ and $\bm{p}_{2}$, and $\phi$ is the angle
between the $\bm{p}_{1}$-$\bm{p}_{2}$ plane and
$\bm{p}_{3}$-$\bm{p}_{4}$ plane.
$\beta_{ij}=p_{j}/(p_{i}^{2}+p_{j}^{2}+2p_{i}p_{j}\cos\theta)^{1/2}$
reduces to $[2 \cos (\theta/2)]^{-1}$ for $i=j$. $\phi_{2}$ is the
azimuthal angle of $\bm{p}_{2}$ with respect to $\bm{p}_{1}$. The
factor $(1+\delta_{ij})^{-1}$ takes into account double counting of
the final states in the case of like particles.

Due to the temperature $T$ we discussed is several
orders of magnitude lower than the nucleonic Fermi temperatures (the
nucleons are strong degenerate), the main contribution to the above
integral comes from the very narrow regions of momentum space near
the corresponding Fermi surfaces $k_F$, just as the calculation of
neutrino emissivity in Ref.~\cite{Y2001}. If the $Z$-factor effect
is included, in the above collision integral, $1-n^{0}(x)$ (and
$n^{0}(x)$) representing the unoccupied (and occupied) state due to
the temperature, should be replaced by
$n(x)|_{T=0}-n(x)=Z_{F}[1-n^{0}(x)]$ (and $Z_{F}n^{0}(x)$) when the
$Z$-factor effect is included.
The collision integral is just attribute to thermal
excitations of particles located in a very narrow region of $\sim
k_B T$ close to their Fermi surfaces, and the state with
$|\epsilon(k)-\epsilon(k_F)|\gg k_B T$ plays no role for the
collision integral because the thermal energy $k_B T$ is too low to
excite those states. Therefore, the high momentum tail makes no
contribution to the collision integral, just as the influence of the
Fermi surface depletion on neutrino emissivity processes discussed
in detail in Ref.~\cite{Dong-SRC2}. Consequently, the collision
integral turns into
\begin{eqnarray}
I_{1i}&=&-\sum_{j}\frac{Z_{Fi}^{2}Z_{Fj}^{2}m_{i}^{*}m_{j}^{*2}k_{B}^{2}T^{2}}{8\pi^{4}\hbar^{6}}\int\int
dx_{2}dx_{3}n^{0}(x_{1})n^{0}(x_{2})\nonumber\\
&\times&[1-n^{0}(x_{3})][1-n^{0}(x_{1}+x_{2}-x_{3})]\int\int
\frac{d\Omega}{4\pi}\frac{d\phi_{2}}{2\pi}\nonumber\\
&\times&\frac{W_{ij}\beta_{ij}}{1+\delta_{ij}}[\psi(\bm{p}_{1})+\psi(\bm{p}_{2})-\psi(\bm{p}_{3})-\psi(\bm{p}_{4})].\label{II}
\end{eqnarray}
Moreover, the driving term of the Landau kinetic equation, which is proportional to $\frac{\partial n}{\partial x}$ at equilibrium state, provides a $Z_{F}$ as well. Therefore, one can include the
$Z$-factor effect in the calculation of the transport coefficients
by adopting $Z_{F}$ both in the collision integral and the driving term by following the derivations in Ref.~\cite{PRB35}. For example, the collision integral reduces to a simple formula of
$I_{1i}=Z_{F}^{4}I_{1i}^{0}$ for pure neutron
matter. One should note that the momentum (energy) flux corresponding to the the shear viscosity (thermal conductivity) also includes $\frac{\partial n}{\partial x}$, Consequently, the shear viscosity (thermal
conductivity) is given by $\eta=\eta^{0}/Z_{F}^{2}$
($\kappa=\kappa^{0}/Z_{F}^{2}$) for pure neutron matter, where $\eta^{0}$ ($\kappa^{0}$) is
the corresponding transport coefficient without the inclusion of the
$Z$-factor effect.


Within the BHF approach, the EOSs of symmetric nuclear matter
($\beta=0$), pure neutron matter ($\beta=1$), and $\beta$-stable
matter, where $\beta=(\rho_{n}-\rho_{p})/(\rho_{n}+\rho_{p})$
denotes the isospin asymmetry with the neutron (proton) number
densities $\rho_n$ ($\rho_p$), are displayed in
Fig.~\ref{fig:EOS}(a). The solid square shows the calculated
saturation point of symmetric matter which is marginally in
agreement with the empirical value due to the introducing of
three-body force. The proton fraction in $\beta$-stable matter is
determined by the density-dependent symmetry energy, i.e., the
isospin-dependent part of the EOS. The EOSs for pure neutron matter
and $\beta$-stable matter show a distinct difference that becomes
more and more visible at high densities, indicating the
non-negligible proton fraction in NS matter. The NS interior is
assumed to be composed of nucleons, electrons and possible muons.
With the conditions of electric neutrality and $\beta$-equilibrium,
the fractions of leptons (electrons and muons as degenerate ideal
gas) and their contributions to the energy density $\varepsilon
(\rho )$ and pressure $p (\rho )$ can be determined uniquely. With
the obtained $\varepsilon (\rho )$ and $p(\rho )$ of the core matter
and the EOS from Baym, Pethick, and Sutherland (BPS)~\cite{BPS} for
crust matter as inputs, the stellar structure, e.g., the density
profile $\rho(r)$ of a static and spherically symmetric NS, is
achieved by solving the Tolman-Oppenheimer-Volkov (TOV) equation.
The established stellar structure is essential for the final
estimation of the shear viscosity timescale and $r$-mode growth time
scale of NSs.

The nucleonic effective mass $m^*$ is defined from the
single-particle energy $\epsilon (p)$ by the relation $m^{\ast
}=k_{F}\left( \partial \epsilon (k)/\partial k\right)
^{-1}|_{k=k_{F}}$. It reduces the
density of states at the Fermi surface with respect to
non-interacting Fermi gas since it is usually smaller than the free
mass. As Ref.~\cite{Baldo2013,shangems}, the rearrangement
contribution of three-body force is not included here. The
calculated effective mass with the BHF approximation are presented
in Fig.~\ref{fig:EOS}(b). The neutron effective mass of pure neutron
matter is not much different from that of $\beta$-stable matter, but
is distinctly larger than that of symmetric matter at the same
density.

\begin{figure}[htbp]
\begin{center}
\includegraphics[width=0.4\textwidth]{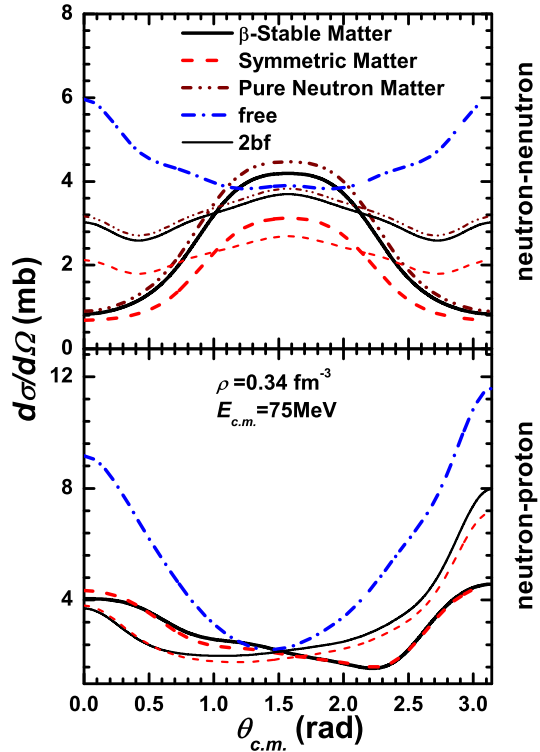}
\caption{(a) Differential cross sections of neutron-neutron
scattering in symmetric matter, pure neutron matter, and
$\beta$-stable matter, taking $\rho=0.34$ fm$^{-3}$ and
center-of-mass energy $E_\text{c.m.}=75$ MeV as an example. (b) The
corresponding total cross sections versus
$E_\text{c.m.}$.}\label{fig:crosssection}
\end{center}
\end{figure}

We calculate the in-medium differential sections within the BHF
method for symmetric matter, pure neutron matter and $\beta$-stable
matter, taking the neutron-neutron scattering at density of $\rho
=0.34$ fm$^{-3}$ (twice the saturation density) as an example, as
shown in Fig.~\ref{fig:crosssection}. The free-space cross section
is also shown for comparison. The in-medium effect leads to a
noticeable suppression of the cross sections, as other calculations
within microscopic nuclear many-body approaches, suggesting the
important role of the medium effect. Our calculated differential
cross sections as functions of center-of-mass scattering angle (and
also the total cross sections versus center-of-mass energy
$E_\text{c.m.}$) have the same shape as that in
Ref.~\cite{Baldo2013} for density $\rho =0.35$ fm$^{-3}$, although
different three-body forces are used. We would like to stress that,
the inclusion of the three-body force increases the cross section at
high $E_\text{c.m.}$, which is in agreement with the conclusion of
Ref. ~\cite{Baldo2013}, but disagrees with the results in
Ref.~\cite{Zhang2012,Zhang2007}.

\begin{figure}[htbp]
\begin{center}
\includegraphics[width=0.45\textwidth]{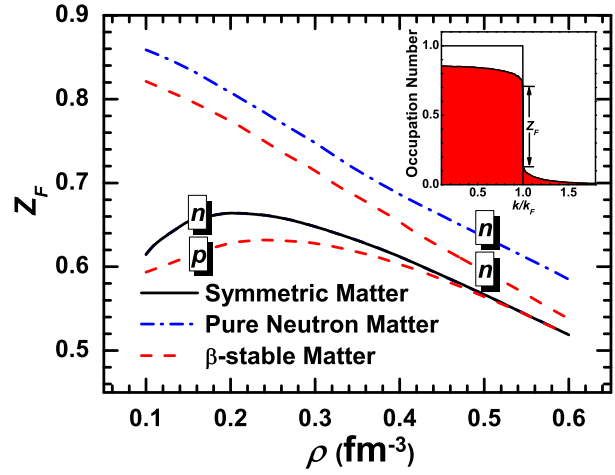}
\caption{Density-dependent $Z$-factor at Fermi surfaces in symmetric
matter, pure neutron matter, and $\beta$-stable matter. The inset
presents a schematic illustration of the Fermi surface depletion
induced by the nucleon-nucleon
correlation.}\label{fig:Z}
\end{center}
\end{figure}

Figure~\ref{fig:Z} exhibits the calculated $Z_F$ at
Fermi surfaces for three different nuclear matter configurations by
employing the Brueckner theory where the self-energy is expanded to
the 2nd-order, i.e., $\Sigma=\Sigma_1+\Sigma_2$. The momentum distribution featured by a
high momentum tail and vacant position below the Fermi surface, is
illustrated in the inset. The behavior of
$Z_{F}$ for symmetric matter is consistent with the result in Refs.
\cite{Dong-SRC2,shangzz}. The $Z$-factor is caused by the short-range
repulsion core and tensor force. The tensor force is dominant at low
densities while the short-range repulsion is dominant at high
densities. The nonmonotonic behavior of $Z_F$ for symmetric matter
and $\beta$-stable matter displayed in Fig.~\ref{fig:Z} is exactly
the results of competition between these two effects, and the $Z_F$
is small both at very low and very high densities. On the other
hand, the $Z_F$ exhibits a strong isospin dependence. At a given
total nucleon density, the $Z_F$ of symmetric matter is smaller
obviously than that of pure neutron matter, that is, the correlation
in the former is stronger than that in the later, because the
$^3SD_1$ tensor interaction component between neutrons and protons
is quite strong in symmetric matter but is completely absent in pure
neutron matter. Namely the pure neutron matter is much closer to the
ideal degenerate Fermi gas, as pointed out in
Ref.~\cite{Science2014}. The results displayed in Fig.~\ref{fig:Z}
will be applied in the following calculations of transport
coefficients.

\begin{figure*}[htbp]
\begin{center}
\includegraphics[width=0.75\textwidth]{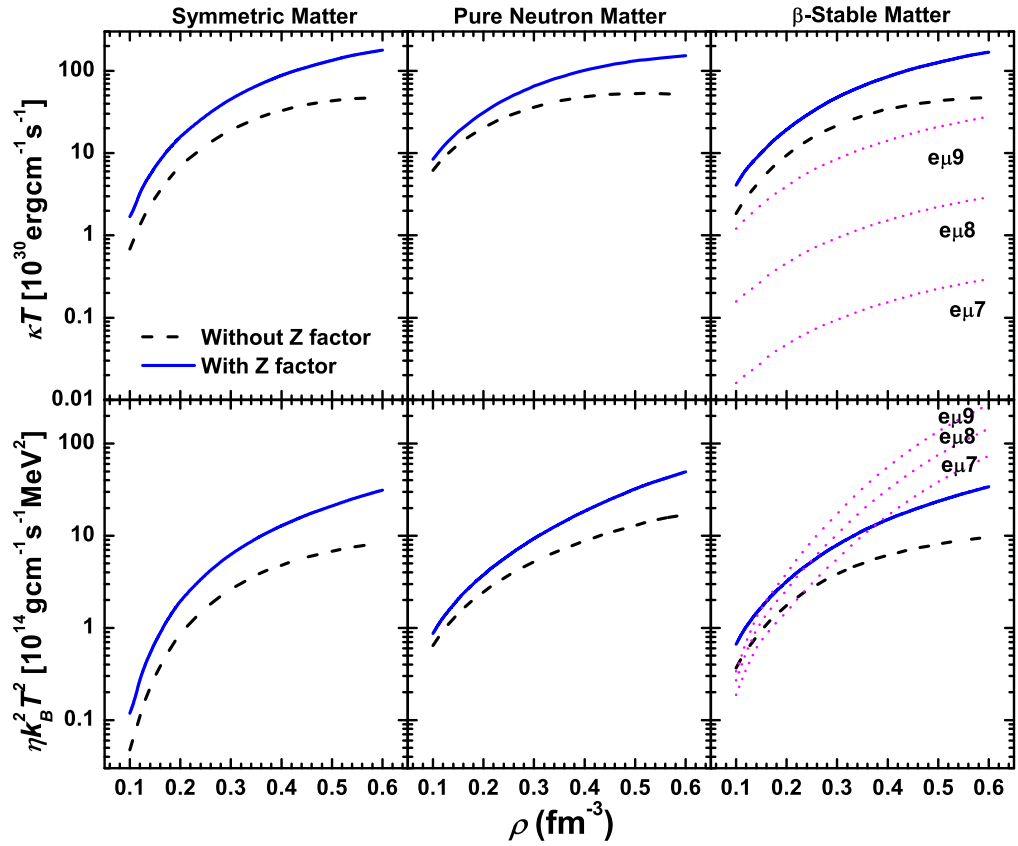}
\caption{Thermal conductivity $\kappa$ (upper panel) and shear
viscosity $\eta$ (lower panel) of nucleons and leptons as a function
of density in symmetric matter, pure neutron matter, and
$\beta$-stable matter. The nucleonic $\kappa_N$ and $\eta_N$ are
calculated with the help of BHF approach without and
with the inclusion of $Z$-factors.}\label{fig:transport}
\end{center}
\end{figure*}

When combining all the results that have been discussed above, we
can now compute the density-dependent shear viscosity under various
temperatures stemming from nucleon-nucleon collisions. The phase
space is quenched in Eq. (2) because of the depletion of Fermi
surface, and therefore the thermal conductivity $\kappa$ and shear
viscosity $\eta$ are increased. The calculated
temperature-independent combinations $\kappa T$ and $\eta T^{2}$
versus density are plotted in Fig.~\ref{fig:transport},
respectively, without and with the inclusion of the $Z$-factor
effect. The lepton (electron and muon) shear viscosity $\eta _{e\mu
}$ and thermal conductivity $\kappa _{e\mu }$ mediated by collisions
of leptons with charged particles in electrically neutral NS matter,
are taken from Ref.~\cite{Shternin2008}. Since the nucleon shear
viscosity $\eta_N$ is mediated by nucleon-nucleon collisions via
strong nuclear force, the $\eta_N$ and $\eta _{e\mu }$ can be
treated independently. Yet, the $\eta _{e\mu }$ ($\kappa _{e\mu }$)
has different temperature-dependent behavior as $\eta_N$ ($\kappa_N
$). So here we show three cases: $T=10^7$, $10^8$, and $10^9$ K. The
relation between $\eta_N$ and $\eta _{e\mu }$ is temperature
dependent, that is, $\eta_N$ becomes more and more important as
temperature decreases. The proton contribution to the shear
viscosity can be neglected safely since the proton contribution is
just $15\%$ even at high density of $\rho=0.6$ fm$^{-3}$.

The $Z$-factor effect enhances the nucleonic $\kappa $ and $\eta$
for the three nuclear matter configurations, in particular at high
densities. For example, at the density of $\rho =0.6$ fm$^{-3}$, the
$\kappa_N$ and $\eta_N$ can be enhanced by about three to four times by the $Z$-factor effect. The nucleonic thermal
conductivity is much larger than the lepton ones for all densities
of NS matter and temperatures of interest. Yet, the situation is
different for shear viscosity. Without the $Z$-factor effect
($Z=1$), the primary contribution to the shear viscosity
$\eta=\eta_N+\eta _{e\mu }$ comes from the lepton scattering which
is just exceeds by nucleon scattering at low densities, in agreement
with the conclusion of Ref.~\cite{Baldo2013}. Once the $Z$-factor is
taken into account, the $\eta_N$ and $\eta _{e\mu }$ become
comparable at intermediate densities, and the $\eta_N$ is about four times larger than $\eta _{e\mu }$ at crust-core
transition density $\rho \approx 0.08$ fm$^{-3}$.

It is widely believed that superfluidity plays a crucial role in NS
dynamics, such as NS cooling and the observed pulsar glitch. It draw
wide attention in communities of nuclear physics and NS physics in
particular after the rapid cooling of the NS in Cassiopeia A was
observed. The strong nuclear force provides several attractive
channels between nucleons in which superfluidity is possible
\cite{sh1,sh2,sh3}. The neutrons dripped out from the neutron-rich
nuclei in NS inner crust, are expected to be paired in a $^1S_0$
singlet state with energy gap of $\sim 1.5$ MeV~\cite{Lombardo2001}.
The proton gas is so dilute that the proton $^1S_0$
superconductivity (superfluidity of charged particles) may survive
until deep inside the star but the neutron $^1S_0$ superfluidity
vanishes because the nuclear interaction in the $^1S_0$ channel
becomes repulsive at short distances for high neutron density.
Nevertheless, at high density, neutron-neutron coupling in the
$^3PF_2$ anisotropic pairing state could appear owing to the
attractive component of the nuclear interaction in this coupling
channel. The coupling between the $^3P_2$ and $^3F_2$ states is
attributed to tensor force. This neutron $^3PF_2$ superfluidity is
of great interest because it was employed to explain the rapid
cooling of the NS in Cassiopeia A~\cite{Page1}. However, the
superfluidity may reduced significantly by the nucleon-nucleon
correlation \cite{Dong-SRC1,Dong-SRC2,shangbcs}.
By performing fittings with several parameters, the
density-dependent gap for the neutron $^3PF_2$ superfluidity of
$\beta$-stable matter is given by~\cite{Dong2020}
\begin{eqnarray}
\Delta _{n}(\rho ) &=&(0.943\rho -0.050)\exp \left[ -\left( \frac{\rho }{%
0.177}\right) ^{1.665}\right] ,
\end{eqnarray}
with a peak value of about 0.04 MeV at $\rho=0.17$ fm$^{-3}$.
The proton $^1S_0$ superfluid gap exists in a
rather narrow region and is much smaller than the neutron $^3PF_2$
superfluid gap as stressed in~\cite{Dong2020}. In addition, the
proton fraction is much smaller than the neutron one for
$\beta$-stable NS matter. Therefore, we do not consider it in the
present work. Here we only focus on the effects of neutron triplet
superfluidity on shear viscosity. As mentioned in
Ref.~\cite{Andersson2005}, we introduce a suppression factor to
estimate the nucleon shear viscosity via
$\eta^{(\text{SF})}_N\approx R_n \eta_N$, where $R_n $ is written
as~\cite{Andersson2005}
\begin{eqnarray}
R_{n} &\simeq &\left[ 0.9543+\sqrt{0.04569^{2}+(0.6971y)^{2}}\right] ^{3} \nonumber \\
&&\cdot \exp \left[ 0.1148-\sqrt{0.1148^{2}+4y^{2}}\right]
\end{eqnarray}
with $y=\Delta (T)/T$. $\Delta (T)$ is the temperature-dependent
energy gap, and the critical temperature is $T_c = 0.57\Delta
(T=0)$. The $\eta_N$ due to neutron-neutron scattering drops
exponentially because of sharp decrease of the number of momentum carriers
near the Fermi surface.

\begin{figure}[htbp]
\begin{center}
\includegraphics[width=0.45\textwidth]{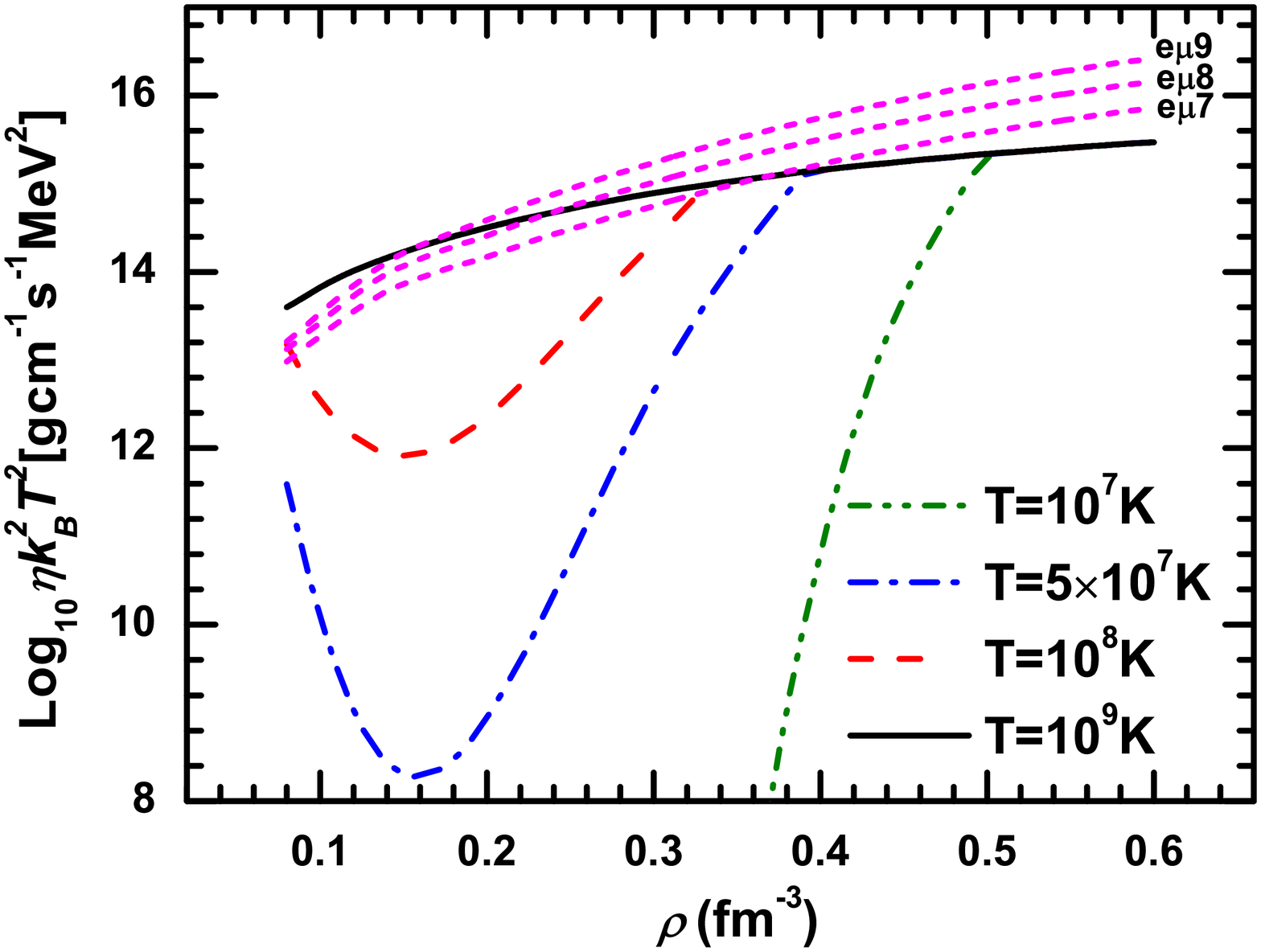}
\caption{Shear viscosity stemming from nucleon-nucleon scattering as
a function of density in $\beta$-stable matter with the inclusion of
neutron triplet superfluidity.}\label{fig:SF}
\end{center}
\end{figure}

The $\eta T^2$ of each component as a function of density under
different temperatures $T$ in the presence of neutron $^3PF_2$
superfluidity are displayed in Fig.~\ref{fig:SF}. If the core
temperatures of NSs are higher than $\sim 2\times 10^8$ K, the
neutron $^3PF_2$ superfluidity disappears. The neutrons in stellar
core becomes superfluid as soon as the NS cools below the critical
temperatures, and accordingly the neutron-neutron scattering is
strongly depressed and the main contribution to the shear viscosity
comes from electron scattering processes. As a result, the
$Z$-factor-quenched superfluid effect plays an opposite role
compared with the $Z$-factor effect itself, and intriguingly it can
be much more significant. For instance, at the temperature $T =
5\times 10^7$ K, the nucleon shear viscosity $\eta_N$ is reduced by
about six orders of magnitude at $\rho=0.17$ fm$^{-3}$, and this
suppression is stronger at lower temperatures. It was concluded in
other references such as~\cite{IV2012} that, at low temperatures $T
< 10^7$K, the contribution to the shear viscosity from the neutron
scattering is more important than the lepton scattering. However,
the $\eta_{e\mu}$ is still larger than $\eta_{N}$ in the presence of
such neutron triplet superfluidity. For example, at temperature $T =
10^7$ K, the $\eta_{N}$ of the nucleon scattering can be neglected
at density $\rho < 0.5$ fm$^{-3}$ in superfluid matter.

\begin{table}[h]
\label{table1} \caption{The calculated shear viscosity time scale
$\tau_\eta$, compared with gravitation-radiation-driven $r$-mode
time scale $\tau_{\text{GW}}=196$ s for canonical neutron stars
rotating at 716 Hz. The results with and without the neutron triplet
superfluidity (SF) are listed, and the weights of the nucleon
contribution are present in the brackets.}
\begin{ruledtabular}
\begin{tabular}{ccccc}
Temperature (K) & $\tau_\eta^{\text{nSF}}$(s)  & $\tau_\eta^{\text{SF}}$(s) \\
\hline
$10^6$ & 402 (66\%) & 1200 (0\%) \\
$10^7$ & $2.99\times10^4$ (50\%) & $4.38\times10^4$ (9\%) \\
$10^8$ & $2.05\times10^6$ (34\%) & $2.26\times10^6$ (27\%) \\
$10^9$ & $1.38\times10^8$ (23\%) & $1.07\times10^8$ (23\%) \\
\hline
\end{tabular}
\end{ruledtabular}
\end{table}

After the stellar structure is established by solving the TOV
equation with the BHF EOS as an input, the time scales of shear
viscosity and of gravitation-radiation-driven growth of $r$-mode for
$1.4M_{\odot }$ canonical NSs are calculated. The overall time scale
is $1/\tau=-1/\tau_{\text{GW}}+1/\tau_{\eta}$, and if
angular-velocity-dependent $\tau_{\text{GW}}$ is smaller than
temperature-dependent $\tau_{\eta}$, the $r$-mode amplitude will
exponentially grow, resulting in $r$-mode instability. The equation
of $1/\tau=0$ determines the critical frequency in
frequency-temperature space, above which is the usually referred to
as the $r$-mode instability window~\cite{Andersson2001,Haskell2015}.

Table I lists the calculated shear viscosity $\tau_{\eta}$ and
$r$-mode growth time scale $\tau_{\text{GW}}$ for canonical NSs. In
non-superfluid NSs, the nucleon-nucleon scattering is indeed the
dominant dissipation mechanism at low temperatures. If the
superfluid effect is included, the situation is completely opposite.
The $\eta_N$ becomes less and less important and even negligible as
temperature decreases. The $\tau_{\eta}$ is enlarged because of the
superfluid effect, indicating weaker shear viscosity damping. It is
generally believed that the $r$-mode instability limits the rotating
angular velocity of accretion millisecond pulsars. At present, the
fastest spinning pulsar is PSR J1748-2446ad spinning at 716
Hz~\cite{PSR716}, and its corresponding $r$-mode growth time scale
$\tau_{\text{GW}}$ is 196 s if $M_{\text{TOV}}=1.4M_{\odot }$ is
assumed. At low temperatures $T=10^6$ K, the shear viscosity
$\tau_{\eta}$ is 402 Hz for nonsuperfluid NS core matter which is
comparable with the $\tau_{\text{GW}}$, and the weight of nucleonic
contribution is as large as $66\%$. However, if the superfluidity is
taken into account, the nucleon-nucleon scattering does not
contribute to the $\tau_{\eta}$ at such low temperature, and the
$\tau_{\eta}$ is much larger than the $\tau_{\text{GW}}$ and hence
the shear viscosity is not much help to damp the $r$-mode
instability. Some authors proposed that the viscous dissipation at
the viscous boundary layer of perfectly rigid crust and fluid core
is the primary damping mechanism. However, it is questioned if the
core-crust boundary is defined by a continuous transition from
non-uniform matter to uniform matter through "nuclear pasta"
phases~\cite{PP1998} and consequently the viscous boundary layer is
smeared out~\cite{Gearheart}.

\begin{figure}[htbp]
\begin{center}
\includegraphics[width=0.45\textwidth]{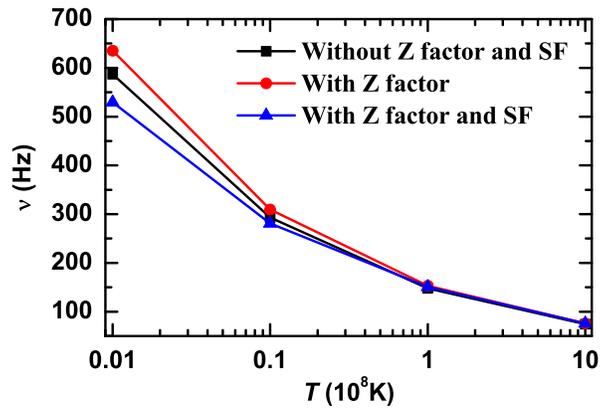}
\caption{The calculated $r$-mode instability critical curves without
the superfluidity (SF) and $Z$-factor, with Z-factor only, with both
the $Z$-factor and neutron triplet superfluidity, are shown for
comparison.}\label{fig:rmode}
\end{center}
\end{figure}

In order to more clearly reveal the roles of the $Z$-factor and
superfluid effects on the $r$-mode instability, the calculated
$r$-mode instability critical curves are presented in
Fig.~\ref{fig:rmode}. The $Z$-factor effect is conducive to damping
the gravitational-wave-driven $r$-mode growth of NSs, in particular
at low temperatures. However, the neutron triplet superfluidity
plays an opposite role and is more significant. At temperatures
higher than $\sim 10^8$ K, both of the two effects are weak, which
is because the neutron-neutron scattering contributes secondary to
shear viscosity and the superfluidity is almost vanishes at such
temperatures. The core temperature of NSs in low mass X-ray binaries
are estimated to be $(1\sim 5) \times 10^8$ K~\cite{Ho2012} and
$10^7\sim 10^8$ K if the direct Urca process opens~\cite{Dong2020},
therefore the shear viscosity cannot be expected to stablize NSs
against $r$-mode oscillations in practical situation. Additional
damping mechanisms perhaps is required.

In summary, the $Z$-factor effects on the thermal conductivity and
shear viscosity have been calculated based on the AK framework,
where the $Z$-factor at Fermi surfaces ($Z_F$), the in-medium cross
sections, nucleon effective masses, and the EOS of NS matter, are
calculated by using the Brueckner theory with the two-body AV18
interaction plus microscopic three-body force. The nucleon-nucleon
correlations, induced by the effects of short-range repulsion and
tensor component of nuclear force, gives rise to the Fermi surface
depletion, i.e., the $Z$-factor effect. The calculated $Z_F$ of
neutrons and protons at Fermi surfaces presents a strong isospin
dependence due to the strong neutron-proton $^3SD_1$ tensor
interaction. The two transport coefficients are enlarged by several
times for symmetric matter, pure neutron
matter and $\beta$-stable matter. The nucleonic thermal conductivity
$\kappa_N $ is much more important than lepton ones for different
densities and temperatures that we considered here, whether or not
this $Z$-factor effect is included. As temperature decreases, the
nucleon shear viscosity $\eta_N$ becomes more and more important
with respect to the lepton contribution $\eta _{e\mu }$. If we take
into account the $Z$-factor effect, the $\eta_N$ may become
comparable with $\eta _{e\mu }$ at intermediate densities, and
larger than $\eta _{e\mu }$ at low densities. As concluded in the
previous works \cite{Dong-SRC1,Dong-SRC2}, the $Z$-factor effect
suppresses the proton $^1S_0$ and neutron $^3PF_2$ superfluidity
strongly, and the proton $^1S_0$ superfluidity almost vanishes.
Contrary to the role of $Z$-factor itself, neutron superfluidity is
able to reduce the shear viscosity significantly (by several orders
of magnitude) when the temperature drops below the critical
temperature. As a result, the contribution to the shear viscosity
from the lepton scattering is still more important than that from
the nucleon scattering at low temperature for the densities of
interest in superfluid matter. Finally, the shear viscosity time
scales $\tau_{\eta}$ along with the time scales $\tau_{\text{GW}}$
of $r$-mode growth due to the emission of gravitational waves for
canonical NSs are calculated. At low temperatures, the
nucleon-nucleon scattering indeed contributes mainly to the shear
viscosity time scale $\tau_{\eta}$. However, if the
$Z$-factor-quenched superfluidity is present, it is less important
and even negligible. In a word, the appearance of superfluidity is
not favorable to damping the $r$-mode instability of NSs. {\color{red}}The
calculated $\tau_{\eta}$ is much larger than the $\tau_{\text{GW}}$
and hence the shear viscosity is not able to damp the $r$-mode
instability even for very cold NSs with core temperature of $10^6$
K. The present work stretches our understanding of the $r$-mode instability of pulsar physics.\\

This work was supported by the National Natural Science Foundation
of China (Grants No. 11775276, 11975282), the Strategic Priority
Research Program of Chinese Academy of Sciences (Grant No.
XDB34000000), the Youth Innovation Promotion Association of Chinese
Academy of Sciences (Grant No. Y201871), the Continuous Basic
Scientific Research Project (Grant No. WDJC-2019-13), the Leading
Innovation Project (Grant No. LC 192209000701), and the Continuous
Basic Scientific Research Project (Grant No. WDJC-2019-13).

\end{document}